\documentclass[10pt,final,letterpaper,twocolumn,conference]{IEEEtran}
\IEEEoverridecommandlockouts
\usepackage{cite}
\usepackage{amsmath,amssymb,amsfonts}
\usepackage{algorithmic}
\usepackage{graphicx}
\usepackage{textcomp}
\usepackage[dvipsnames]{xcolor}

\def\BibTeX{{\rm B\kern-.05em{\sc i\kern-.025em b}\kern-.08em
    T\kern-.1667em\lower.7ex\hbox{E}\kern-.125emX}}
    
\newcommand{\honeymodels}{\texttt{HoneyModels}}
\newcommand{\honeymodel}{\texttt{HoneyModel}}

\newif{\iffinal}    
\finaltrue

\newif{\ifmilcomfinal}
\milcomfinalfalse
\usepackage[ruled,linesnumbered]{algorithm2e}    
\usepackage{amsmath}                             
\usepackage{amssymb}                             %
\usepackage{booktabs}                            
\usepackage{graphicx}                            
\usepackage{lipsum}                              
\usepackage{mathtools}                           
\usepackage[all]{nowidow}                        
\usepackage{optidef}                             
\usepackage{pgfplotstable}                       
\usepackage{subcaption}                          
\usepackage{siunitx}                             
\usepackage{substr}                              
\usepackage{float,graphicx}                      %
\usepackage{calrsfs}
\usepackage{lipsum}
\usepackage{setspace}
\usepackage{tabto}
\usepackage{multirow}
\usepackage{adjustbox}
\usepackage{comment}
\usepackage{svg}
\usepackage{pgf}
\usepackage[ruled,linesnumbered]{algorithm2e}    

\DeclareMathAlphabet{\pazocal}{OMS}{zplm}{m}{n}
\DeclarePairedDelimiter{\norm}{\lVert}{\rVert}

\newcommand{\ie}{i.e.,}                          
\newcommand{\eg}{e.g.,}                          

\makeatletter

\usetikzlibrary{plotmarks}
\newcommand\marksymbol[2]{\tikz[#2,scale=1.2]\pgfuseplotmark{#1};}

\ifmilcomfinal
\usepackage[draft]{hyperref}
\else
\usepackage[hidelinks,bookmarks=False]{hyperref}
\fi

\begin{document}

\title{HoneyModels: Machine Learning Honeypots\\
}

\iffinal
    \author{\IEEEauthorblockN{Ahmed Abdou\IEEEauthorrefmark{1}, Ryan Sheatsley\IEEEauthorrefmark{1}, Yohan Beugin\IEEEauthorrefmark{1}, Tyler Shipp\IEEEauthorrefmark{2} and Patrick McDaniel\IEEEauthorrefmark{1}}
    \IEEEauthorblockA{\IEEEauthorrefmark{1}School of Electrical and Computer Engineering\\
    The Pennsylvania State University, University Park, USA\\
    Emails: aabdou@psu.edu sheatsley@psu.edu ybeugin@psu.edu mcdaniel@cse.psu.edu}
    \IEEEauthorblockA{\IEEEauthorrefmark{2}DEVCOM C5ISR, Maryland, USA\\
    Email: tyler.j.shipp3.ctr@mail.mil}}
\else 
    \author{Anonymous}
\fi

\maketitle

\begin{abstract} 
\label{abstract}
Machine Learning is becoming a pivotal aspect of many systems today, offering newfound performance on classification and prediction tasks, but this rapid integration also comes with new unforeseen vulnerabilities. To harden these systems the ever-growing field of Adversarial Machine Learning has proposed new attack and defense mechanisms. However, a great asymmetry exists as these defensive methods can only provide security to certain models and lack scalability, computational efficiency, and practicality due to overly restrictive constraints. Moreover, newly introduced attacks can easily bypass defensive strategies by making subtle alterations. In this paper, we study an alternate approach inspired by honeypots to detect adversaries. Our approach yields learned models with an embedded watermark. When an adversary initiates an interaction with our model, attacks are encouraged to add this predetermined watermark stimulating detection of adversarial examples. We show that \honeymodels{} can reveal $69.5$\% of adversaries attempting to attack a Neural Network while preserving the original functionality of the model. \honeymodels{} offer an alternate direction to secure Machine Learning that slightly affects the accuracy while encouraging the creation of watermarked adversarial samples detectable by the \honeymodel{} but indistinguishable from others for the adversary.
\end{abstract}


\section{Introduction} 
\label{introduction}

Machine Learning is quickly overtaking traditional programming methods and providing systems the ability to make accurate and swift predictions and classifications. This paradigm shift can be observed today in self-driving cars \cite{bojarski2016end}, financial transactions \cite{bose2001business}, or medical imaging \cite{wernick2010machine}. The ubiquity of Machine Learning systems and their integration into various fields pose a serious security threat, as these systems have been shown to be inept to carefully crafted adversarial inputs \cite{szegedy2013intriguing}. Real-life examples of these failures have been demonstrated and they include, among many others, glasses that can deceive face-recognition systems \cite{sharif2016accessorize} or adversarial examples to fool systems operating in the physical world \cite{kurakin2017adversarial}. The feasibility of these attacks have been expanded by research in black-box settings, \ie{} techniques that require only the ability to query the model and no additional knowledge about the model to craft adversarial examples \cite{papernot2017practical}.

\begin{figure}[ht]
\centering
\includegraphics[width=\columnwidth]{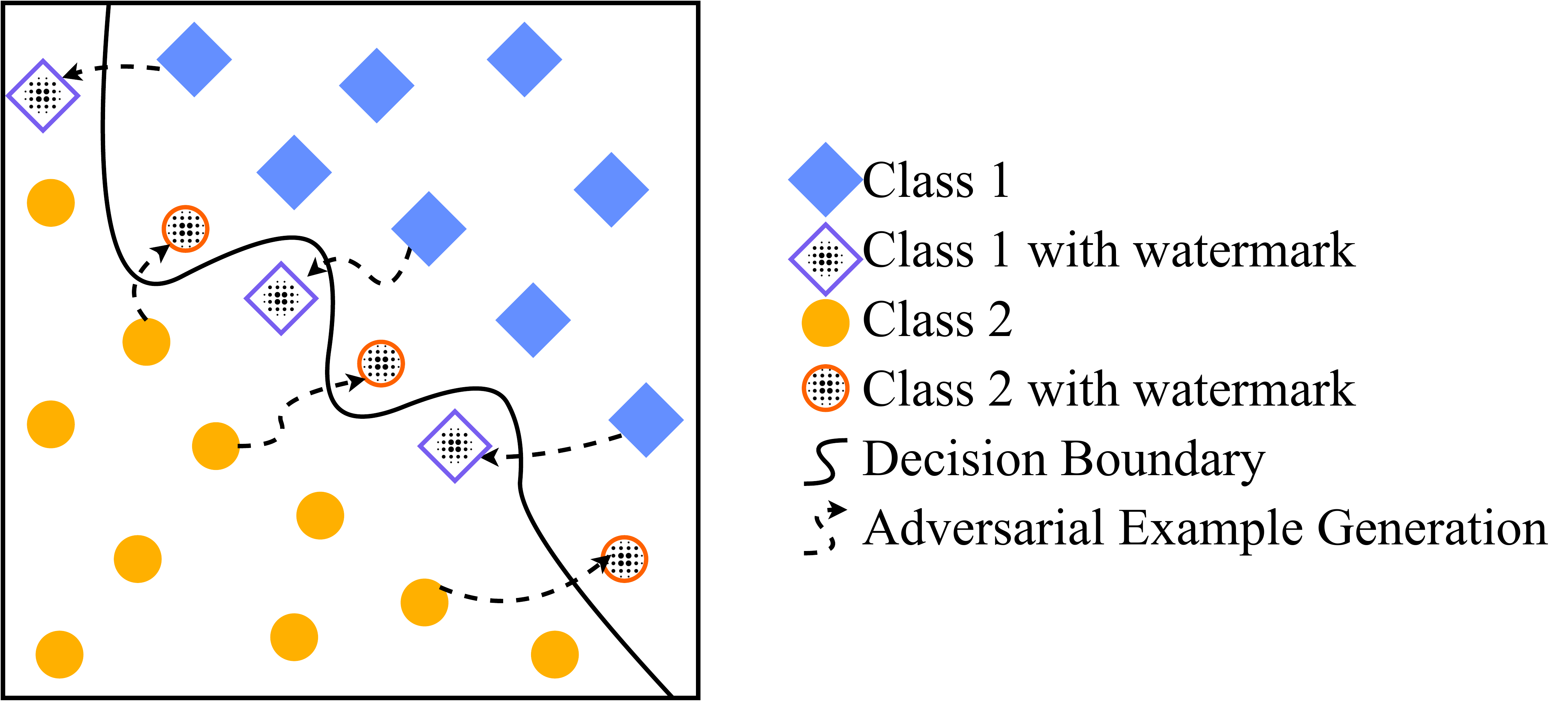}
\caption{Illustration of the \honeymodels{} mechanism on a binary classifier; poisoned samples within the training data make adversaries converge towards watermarked examples.}
\label{fig:illustration}
\end{figure}

These vulnerabilities of Neural Networks were shown to be transferable to simpler models with varying characteristics and parameters \cite{papernot2016transferability}. Concerned with the latter, a period of rapid growth within the field has initiated. Defenses such as Distillation \cite{papernot2016distillation} and many others have been shown to fail against new first order attack methods \cite{carlini2017towards}. Consequently new lines of work have attempted to provide robustness for Neural Networks. Some of the most notable defenses are based on adversarial training \cite{madry2017towards}, randomized smoothing \cite{cohen2019certified}, semi-definite programming \cite{raghunathan2018certified}, mixed-integer programming \cite{tjeng2017evaluating}, and linear programming \cite{wang2018mixtrain}. While successful, these methods are limited by high-computation times, scalability issues, and restrictions to certain datasets or architectures making them infeasible for practical deployment. 

Inspired by honeypots and the inherent need of a defensive system in Machine Learning that is not overly restrictive, computationally inexpensive, and scalable; we offer an alternate approach to secure deep learning systems. Our approach securely creates a secret key that is used to alter the values of features inside a model and create a unique watermark. This watermark is reproduced by any adversary mounting a gradient based attack against our model (see \autoref{fig:illustration}). Detection mechanisms are used inference-time to detect these alterations and predict if a sample is adversarial with high-probability. Model weights are adjusted in the training procedure through self-poisoning to yield a model that functions correctly while leading attack gradients to reproduce the predetermined watermark. The watermark preserves accuracy and semantics of the model while being indistinguishable to adversaries when reconstructed and analyzed. \honeymodels{} works by exploiting the aggressive nature of the adversary attempting to find a small perturbation that causes misclassification.

We evaluate \honeymodels{} abilities to have gradient based attacks prioritize watermarked features and to detect adversarial inputs along three axes. These criteria are \textbf{Viability}: the creation and deployment of a \honeymodel{} should not affect its accuracy. \textbf{Detection}: \honeymodels{} should detect a high-percentage of adversaries while limiting false-positives. \textbf{Indistinguishability}: adversarial samples generated from a \honeymodel{} and a benign model should be statistically similar such that an adversary can not distinguish them from one another. These criteria were evaluated on Neural Network Classifiers with two image datasets: MNIST and CIFAR10. The detection of adversarial inputs crafted by state-of-the art attack algorithms in a white-box setting was evaluated and we show that we are able to successfully detect up to $69.5$\% of adversaries while having a false positive rate of $14.1$\%. This can be accomplished through slight alterations to the training process which have little effect on the models accuracy, training time, and semantics.

Mainly our contributions can be summarized as follows: 
\begin{itemize}
 \item We propose a solution to secure Neural Networks that involves detection of adversarial inputs through adjustment of model weight that induce the predetermined watermark. 
 \item We evaluate the effectiveness of \honeymodels{} to reveal adversaries on popular datasets of varying size and type and further relay our efficacy through evaluating against adaptive adversaries.
 \item We show that an adversary cannot distinguish between adversarial examples generated by a \honeymodel{} and a regular model using probability distribution metrics. This prevents adversaries from deducing the watermark.
\end{itemize}
\section{Background} 
\label{background}
In this section we characterize the vulnerability of Neural Networks to adversarial examples and the attack algorithms that have culminated as a result. The inner-workings of these algorithms allow us to exploit them and create a technique to detect adversaries. We also explore the techniques used to embed watermarks within Machine Learning models.

\subsection{Crafting an Adversarial Example}
An adversary can create an adversarial example to fool a Neural Network. To do so, an adversary begins with some baseline input $x$ and knowledge of the model which they aim to attack (\eg{} model architecture, training and test datasets, output probabilities, etc.). Attacks that assume no previous knowledge about the model are called black-box attacks, while those that do have knowledge are classified as white box attacks \cite{papernot2017practical}. The goal of the adversary is to find some minimal noise $\delta$ such that when added to the baseline input $x$ creates some $\hat{x}$ close to $x$ (norms $\norm{.}$ are typically used to quantify this proximity) that is misclassified by the model. We can formalize the space of adversarial examples around $x$ as the ball $B_{\delta}(x) \triangleq \{ \hat{x} ~|~ \norm{ \hat{x} - x} \leq \delta \}$ where $\delta \geq 0$ \cite{szegedy2013intriguing}. 

The attack algorithms are greedy in nature prioritizing noise that greatly affects the classification of an input. Gradient-based attacks lower-bound this attack problem in various ways and have a different notion of minimal noise. For example the Jacobian Saliency Map Attack (JSMA) \cite{papernot2016limitations} counts the finite number of altered features, while the Carlini-Wagner (CW) attack \cite{carlini2017towards} uses some norm to measure the minimal amount of noise required to cause a misclassification. 

\subsection{Watermarked Models}

The process of embedding watermarks in a model has been previously explored to mitigate model theft. Watermarked models are usually triggered at inference time with specific inputs; Deepsigns \cite{darvish2019deepsigns} observes the activation for these inputs while others \cite{adi2018turning} will use misclassification to guarantee ownership. \honeymodels{} embedding process similarly exploits the over-parameterization of Machine Learning models \cite{gu2017badnets}.

\begin{figure*}[t]
\centering
\includegraphics[width=0.8\textwidth,height=3cm]{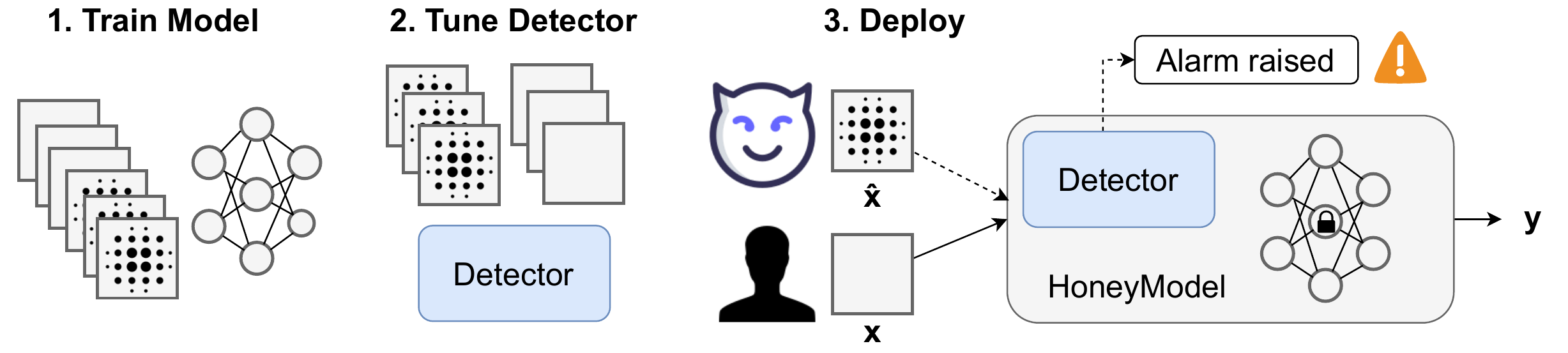}
\caption{The pipeline for creating a \honeymodel{} can be described in three steps: 1. Training the model on the dataset altered with the watermark. 2. Tuning the detector to identify adversarial examples generated on the model. 3. Deploying the \honeymodel{} and punishing adversaries who attempt to attack it.}
\label{fig:pipeline}
\end{figure*}

\section{Methodology}

In this section we begin by introducing the Threat Model which encompasses the assumed adversarial capabilities, objective, and knowledge as they pertain to \honeymodels{}. We then discuss how to create a defensive model that has adversaries reveal themselves when they interact with it. We define the necessary criteria for creating a \honeymodel{} such that it preserves test time accuracy while limiting false positives and is still discrete to adversaries. Explicitly, we present various techniques for creating a secret key, embedding it into a watermark within our model through self-poisoning, and detecting adversarial examples with altered features reproducing the watermark. We provide an easy to implement framework that adheres to the constraints of being highly-scalable and computationally efficient.

\subsection{Threat Model}

The objective of the adversary is to have the Neural Network misclassify an input presented at inference time. This is achieved through interacting with our model and generating an adversarial example that is perceptually similar to the original input. We assume that the adversary has the ability to query and interact with our model in a grey-box setting. The adversaries have full knowledge of model parameters such as weights, architecture, test and training data, and access to respective input probabilities to strengthen the attack. However, they will not know whether our model is a \honeymodel{} or not and will not have access to the secret key used to alter the model and create a unique watermark. When adversaries are attempting to analyze the differences between \honeymodel{} adversarial examples and baseline adversarial samples, they are bounded by the amount of samples they can reproduce,  they are bound to statistical techniques to distinguish between \honeymodel{} and baseline adversarial examples. The security of a \honeymodel{} arises from the Kirckhoffs's principle, meaning that an adversary must know the secret key to know where not to add noise and thus not reproduce the watermark.

\subsection{Criteria}

To assess the efficacy of \honeymodels{}, we evaluate the impact of the introduction of the watermark according to the following criteria: viability, detection, and indistinguishability which are further explained in this section.

\subsubsection{Viability}

The embedding process of the watermark should not greatly affect the model's classification or predictive accuracy. We measure the effect of the size and amplitude of the watermark on the test time classification accuracy. Since embedding the watermark into the model requires self-poisoning, we also measure the model's resiliency to label-poisoning.

\subsubsection{Detection}

When an adversary interacts with a \honeymodel{} to generate an adversarial example, the watermark should be reconstructed successfully with high probability. We measure the ability to detect adversarial samples through various mechanisms in our evaluation.

\subsubsection{Indistinguishability}
The reconstruction of the watermark should be significant for detection by the \honeymodel{}. However, the adversary should not be able to distinguish between watermarked and regular samples. As a result, the adversary would not be able to detect the presence of a watermark, remove it, or craft an adversarial sample that evades it. We measure indistinguishability in our evaluation and explain how it is attained.

\subsection{Creating a \honeymodel{}}

The pipeline for creating a \honeymodel{} can be described in three steps (see \autoref{fig:pipeline}): 1. Training the model on the dataset altered with the watermark. 2. Tuning the detector to identify adversarial examples generated by interaction with the model. 3. Deploying the \honeymodel{} and punishing adversaries who attempt to attack it. Specifically, we explain the process of training a \honeymodel{} which requires the user to both create a watermark and embed it. We elaborate on the new hyper parameters involved with this process and the properties it must follow to preserve its functionality and practicality.

\subsubsection{Selecting Watermark Features}

\begin{figure}[h]
\centering
 \includegraphics[width=.7\columnwidth]{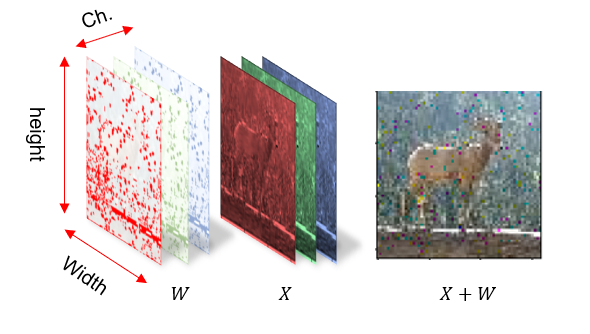}
\caption{Example of a randomly generated watermark.}
\label{fig:watermark}
\end{figure}

The first process of creating a \honeymodel{} involves creating a watermark that we aim to detect at inference time to reveal adversaries. A secret key is generated which describes which features of our training data should be altered and how. Two parameters arise from this key: \textit{Watermak Size} which defines the number of features to be altered and \textit{Amplitude} which describes the amount of alteration to apply. 
The alteration of features should still be semantically preserving. This secret key should be generated in a cryptographic-randomly sense such that an adversary cannot guess it.
In the case of image data, constraints to the key's semantics can be neglected as long as the perturbation introduced by the watermark is not too large and does not completely change the image. In the evaluation, we measure the effects of varying these new parameters (see \autoref{fig:watermark}). 

\subsubsection{Embedding a Watermark through Data Modification}

The watermark is embedded into the \honeymodel{} by applying it to some training samples and changing their labels to misclassify them. As many attacks rely on gradients to compute adversarial examples, our label-flipping scheme encourages the model parameters to, when differentiated, point towards our watermarked examples that lie close to the decision boundary (see \autoref{fig:illustration}).

The watermark is randomly applied to a portion of the training data according to a poisoning percentage. The label of the corresponding poisoned data is flipped to any other different label not equal to the original one. This process is done randomly so that each class of the datasets is equally poisoned, \ie{} for MNIST we should have the same number of mislabeled classes $0-9$. Intuitively, training a model on these modified examples yields a model that when interacted with has gradients that maximizes marginal loss (examples in training data with different labels) between two class pairs while minimizing the distance between the class pairs (mislabeled example has our predetermined watermark). 

\subsubsection{Validation of Watermark Reconstruction}

During or after the training process, the success of the embedding process can be measured through a cosine similarity measure. To do so, we create adversarial samples using a gradient-based attack method, we then extract from each sample the watermark $\hat{W}$ specified by the secret key, and compare it to the original watermark $W$:
\[ Sim(W,\hat{W}) \triangleq \frac{W \cdot \hat{W}}{\norm{W} \norm{\hat{W}}} \]

\subsubsection{Detection of the Watermark}

At inference time, when an input is presented to a \honeymodel{}, the features $\hat{W}$ specified by the secret key, used to embed the watermark at training time, are extracted and compared to the original watermark $W$.  We propose using a Machine Learning detector such as logistic regression to perform this comparison. The process of tuning our detector begins with the generation of a large amount of adversarial samples from different attack techniques. The watermarks are extracted from these adversarial samples and they are used to train a classifier or determine parameters for optimal detection.

\section{Evaluation}

In this section, we elaborate on the datasets used, experimental apparatus, and metrify the viability, detection capability, and indistinguishability of resulting adversarial samples with respect to varying hyper-parameters. 

The details about our experimental setup and the parameters used are available in \autoref{appendix}. 

Mainly we are concerned with answering the following questions:

\begin{itemize}
 \item How does altering the poison percentage and the secret key parameters affect viability, detection, and indistinguishability?
 \item How resilient are models to our label-poisoning technique?
 \item Are \honeymodels{} a good detection mechanism for detecting misuse of Machine Learning models?
 \item Are adversarial samples generated from a \honeymodel{} and benign model indistinguishable?
\end{itemize}

\subsection{Viability}

\begin{figure*}[t]
 \centering
 \begin{subfigure}[t]{\columnwidth}
 \centering
 \resizebox{0.9\columnwidth}{!}{\input{figures/MNIST/MnistP.pgf}}
 \caption{ --- Impact of the poisoning percentage on the accuracy of the MNIST model and the reconstruction accuracy of the watermark for the aforementioned attacks. The watermark size is fixed to $25$\%.}
 \end{subfigure}
 \hfill
 \begin{subfigure}[t]{\columnwidth}
 \centering
 \resizebox{0.84\columnwidth}{!}{\input{figures/MNIST/MnistK.pgf}}
 \caption{ --- Impact of the watermark size on the accuracy of the MNIST model and the reconstruction accuracy of the watermark for the aforementioned attacks. The poisoning percentage is fixed to $40$\%.}
 \end{subfigure}
 
 \medskip
 \begin{subfigure}[t]{\columnwidth}
 \centering
 \resizebox{0.9\columnwidth}{!}{\input{figures/CIFAR10/Cifar10P.pgf}}
 \caption{ --- Impact of the poisoning percentage on the accuracy of the CIFAR10 model and the reconstruction accuracy of the watermark for the aforementioned attacks. The watermark size is fixed to $25$\%.}
 \end{subfigure}
 \hfill
 \begin{subfigure}[t]{\columnwidth}
 \centering
 \resizebox{0.9\columnwidth}{!}{\input{figures/CIFAR10/Cifar10K.pgf}}
 \caption{ --- Impact of the watermark size on the accuracy of the CIFAR10 model and the reconstruction accuracy of the watermark for the aforementioned attacks. The poisoning percentage is fixed to $20$\%.}
 \end{subfigure}
 
 \caption{Parameter scan measuring: {\color{Blue} \textbf{blue}} \protect\marksymbol{triangle*}{Blue} corresponds to the base model, {\color{ForestGreen} \textbf{green}} \protect\marksymbol{*}{ForestGreen} to JSMA, {\color{BrickRed} \textbf{red}} \protect\marksymbol{diamond*}{BrickRed} to PGD, {\color{Purple} \textbf{purple}} \protect\marksymbol{square*}{Purple} to CW.}
 \label{fig:results}
\end{figure*}

To evaluate the viability of \honeymodels{}, we study the impact of the different parameters involved in the watermark embedding on the model accuracy and reconstruction accuracy of the watermark. These parameters are: percentage of features altered, amplitude of alteration for each feature, and data poisoning percentage. Our initial results showed that the amplitude of alteration had to be set to any large value such that the model prioritizes the embedding of these features. Therefore, we fixed the amplitude of alteration to the highest value ($1.0$) and study only the impact of the data poisoning and alteration percentages. To measure the percentage of the original watermark that is reconstructed, we generate adversarial samples on our \honeymodel{} using the CW, PGD, and JSMA attacks and compute the previously mentioned cosine similarity score. 

On \autoref{fig:results}, it can be seen that the accuracy of the base model is gradually degrading with the poisoning percentage. However, the MNIST dataset exhibits a unique phenomena in that it is resilient to label poisoning until about 90\% of data is poisoned where accuracy begins to drop significantly. The reconstruction scores for each adversarial attack do not vary much and the PGD attack reconstructs the watermark the most. This is expected since the PGD attack gradually adds noise to the whole feature space and clips to the respective epsilon value. The JSMA attack comes second in terms of reconstruction since it is bounded by the number of features it can change but maximally changes these features. Finally, the CW attack has the lowest reconstruction score since it aims to find the minimal amount of noise to cause a misclassification. Although the reconstruction percentages for CW, and JSMA are less than 50\% they are still detectable by our detection mechanism.

We show that our models are resilient to the data poisoning method and discovered that we can poison up to 90\% of MNIST data while sacrificing only 12\% of our accuracy. Similarly, we can gradually increase the percentage of features altered while having little to no effect on the accuracy. As a result, our embedding technique is viable as it minimally affects the original functionality of the model.

\subsection{Detection Mechanisms}

\begin{table*}[t]
\begin{adjustbox}{max width =\textwidth}
\begin{tabular}{@{}cccclccclccclccc@{}}
\toprule
\multicolumn{1}{l}{} & \multicolumn{3}{c}{CW} & \multirow{2}{*}{} & \multicolumn{3}{c}{PGD} & \multirow{2}{*}{} & \multicolumn{3}{c}{JSMA} & \multirow{2}{*}{} & \multicolumn{3}{c}{Overall} \\ \cmidrule(lr){2-4} \cmidrule(lr){6-8} \cmidrule(lr){10-12} \cmidrule(l){14-16} 
 & ACC. & FPR & TPR & & ACC. & FPR & TPR & & ACC. & FPR. & TPR. & & ACC. & FPR. & TPR. \\ \midrule
MNIST & 45.0\% & 56.5\% & 97.5\% & \multirow{3}{*}{} & 97.5\% & 4.9\% & 100\% & \multirow{3}{*}{} & 78.5\% & 20.9\% & 77.9\% & \multirow{3}{*}{} & 69.5\% & 14.1\% & 54.0\% \\
CIFAR10 & 21.0\% & 76.6\% & 18.9\% & & 55.5\% & 28.6\% & 40.2\% & & 30.0\% & 70.1\% & 30.7\% & & 56.3\% & 37.8\% & 50.5\% \\ \bottomrule
\end{tabular}
\end{adjustbox}
\caption{Detection accuracy (ACC.) and false (FPR) and true (TPR) positive rates of a logistic regression
classifier trained on both adversarial samples from a \honeymodel{} and benign model.}
\label{tab:results}
\end{table*}

We now evaluate the ability of the detector to reveal adversaries. For that, we trained a classifier on benign and adversarial examples generated by the CW, JSMA, and PGD attacks. Thus, we are able to predict if a sample presented inference time is adversarial or not based on checking the values of the watermark. \autoref{tab:results} presents the results for an optimal percentage of features altered (10\%) and poisoning percentage (30\%). A total of 1,500 adversarial samples, 500 for each attack, were used to train a logistic regression classifier which can detect up to 69.5\% of adversaries attempting to deceive a \honeymodel{}. We remark that PGD is the most detectable adversarial attack. This is because PGD adds noise to the complete input resulting in a larger reconstruction of the watermark. JSMA comes in second because a maximal amount of features in the input are altered completely. Finally, the CW attack is detected 45\% of the time because it is meant to be discrete; this algorithm has a term in its optimization objective that penalizes the total noise added to the input.

\subsection{Indistinguishability of the Resulting Samples}

\begin{figure*}[t]
    \centering
    \resizebox{1.3\columnwidth}{!}{\includegraphics[]{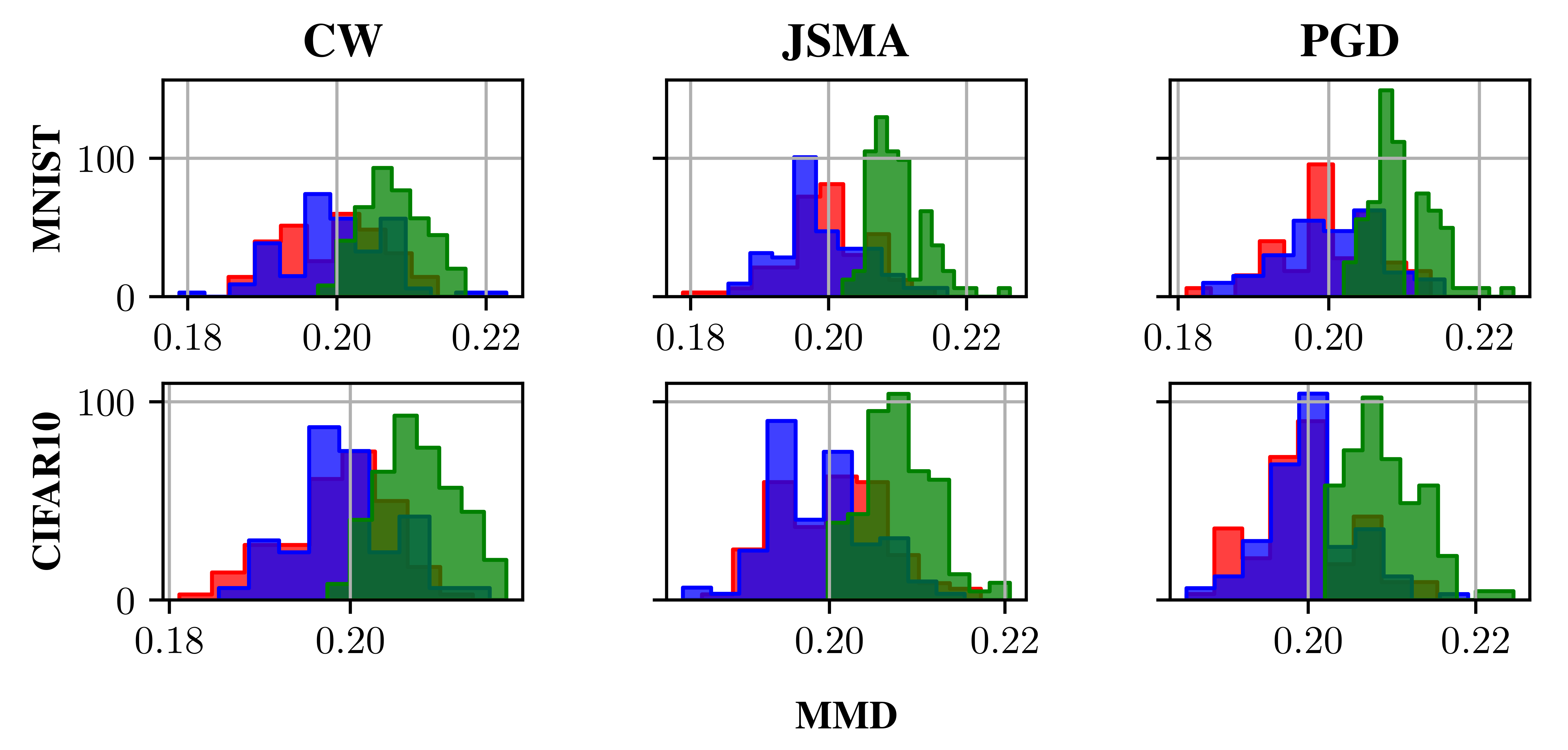}}
    \caption{Distribution of Maximum Mean Discrepancy scores from adversarial samples from a benign model and \honeymodel{} compared among themselves ({\color{Blue} \textbf{blue}} corresponds to samples from the benign model only and {\color{Red} \textbf{red}} to samples from the \honeymodel{} only) and to one another ({\color{ForestGreen} \textbf{green}} is for samples from the benign model and \honeymodel{}).}
    \label{fig:results_indistinguishability}
\end{figure*}

In this section, we quantify the distinction between adversarial samples generated with regular models vs. \honeymodels{}. To measure indistinguishability, we use the Maximum Mean Discrepancy (MMD). The MMD finds the maximal average difference between two distributions $X$ and $Y$ as measured by the kernel functions \footnote{Here, we only used the Gaussian kernel.} in $F$:

\[ \text{MMD} [F,X,Y] \triangleq \sup_{f \in F} \left ( \frac{1}{n} \sum_{i=1}^n f(x_i) - \frac{1}{m} \sum_{i=1}^m f(y_i)\right ) \]

We use bootstrapping to estimate the distribution of MMD values between adversarial samples generated from a \honeymodel{} and a benign model as well as between themselves. We obtain a baseline for the statistical differences of adversarial samples as they relate to themselves through the MMD. If an overlap occurs between these distributions and the benign vs \honeymodel{} distributions we achieve some probability of indistinguishability dependent on the amount of overlap. Note that this technique for measuring distinguishability assumes the most powerful adversary, but in a grey-box setting adversary would be limited by the amount of queries available and may not be able to realize the distribution of MMD scores. 

In \autoref{fig:results_indistinguishability}, we sub-sample 50 adversarial samples for each distribution, compute the MMD scores among each distribution ({\color{Blue} \textbf{blue}} corresponds to samples from the benign model only and {\color{Red} \textbf{red}} to samples from the \honeymodel{} only) and between each other ({\color{ForestGreen} \textbf{green}} is for samples from the benign model and \honeymodel{}), and repeat it 1,000 times with replacement. Distinguishability would occur if an adversary can pick some MMD threshold to separate the green distribution from the blue and red distributions. In most cases, we can see on \autoref{fig:results_indistinguishability} that all three attacks generate \honeymodel{} adversarial samples that are indistinguishable from regular adversarial samples (indicated from a large overlap in the distributions). 
\ifmilcomfinal
\else
\section{Related Work}

In this section, we describe existing mechanisms for detecting adversarial examples during inference time, we describe their shortcomings and feasibility. \honeymodels{} aim to use a detection method that is computationally inexpensive and robust. However, we aim to induce the desired watermark and detect adversarial examples as opposed to just detecting adversarial examples.

\subsection{Statistical-Based Detectors}

Such detectors statistically quantify the distinction between the two distributions of adversarial and regular examples. Examples of these measures include Maximum-Mean Discrepancy (MMD) which is a kernel-based test \cite{gretton2012kernel} or Energy-Distance; a specific case of MMD where no Kernel is used \cite{szekely2013energy}. These methods can not be utilized at inference time when a single example is known and not the complete distribution. However, various detectors have been proposed that utilize Machine Learning classifiers to distinguish between adversarial examples and benign examples to address this problem. Grosse et al. \cite{grosse2017statistical} proposed adding additional classes for classification of adversarial examples which can be viewed as a variant of adversarial training. Similarly, Gong et al. \cite{gong2017adversarial} trained a binary classifier to separate benign examples from adversarial examples. However, stronger attacks were shown to bypass these classifiers.

Recently a new direction in statistical detection quantifies how feature-representations and log-odds change under noise which can even be corrected inference time \cite{roth2019odds}. This method can be combined with a trained logistic classifier that can improve classification accuracy while being robust to adversarial attacks due to low dimensionality, and shallowness. This criteria was shown to be insufficient as an adversary can add an additional loss term in a targeted attack that samples random noise vectors producing an adversarial example that is deep within a decision boundary \cite{hu2019new}. Based on this attack assumption the author proposes calculating the distance of an example from the decision boundary as natural images tend to lie close to decision boundaries \cite{hu2019new}.

\subsection{Feature Squeezing}

Feature Squeezing is based on prediction inconsistency; a detection method that measures the discrepancy of a classification between various models during inference. First, some features of the evaluated input are filtered out by using different methods such as depth-reduction \cite{xu2017feature} or scalar quantization \cite{liang2017detecting}. Then, several models are queried to measure their output disagreement and detect adversarial examples. However, it was later shown that Feature-Squeezing techniques can be fooled by modifying CW and PGD attacks \cite{sharma2018bypassing}.

\subsection{Trapdoor}
Trapdoor functions seek to guide adversarial crafting algorithms towards particular inputs that are known to the defender to be produced by adversaries. \cite{shan_gotta_2020} introduces a trapdoor method, where attack algorithms reproduce particular inputs embedded on the classification manifold. To verify whether or not an input is a trapdoor, the input neuron activations are compared to trapdoor neuron activations computed offline. If they are sufficiently similar, then a flag is raised. Our work on \honeymodels{} is inspired by the work by~\cite{shan_gotta_2020}, yet we provide a different mechanism for embedding the trapdoor (i.e., our watermark), a machine learning-agonistic approach to detection (i.e., MMD), and extend our work to also include an anaylsis on the \textit{indistinguishability} of the watermark (of which \cite{shan_gotta_2020} defers to future work). To this end \cite{shan_gotta_2020} provides a complementary approach to \honeymodels{}, but it is unclear if the produced trapdoors are detectable by the adversary. 
\fi
\section{Conclusion} 

In this work, we proposed \honeymodels{}: Machine Learning Honeypots that detect adversarial usage of Machine Learning models. Our technique works by securely generating a key that specifies features to alter, creating a unique watermark that is embedded within our model through exploiting label poisoning, and training a detector to detect adversarial samples crafted against our model. We showed on Neural Networks with different datasets of varying dimensionality that adversarial samples generated from a \honeymodel{} are indistinguishable to those of a benign model and therefore an adversary can not tamper with the embedded watermark. \honeymodels{} offer an alternate approach to secure Machine Learning models against adversarial examples.
\section*{Acknowledgments}
\iffinal
This research was sponsored by the U.S. Army Combat Capabilities Development Command Army Research Laboratory and was accomplished under Cooperative Agreement Number W911NF-13-2-0045 (ARL Cyber Security CRA). The views and conclusions contained in this document are those of the authors and should not be interpreted as representing the official policies, either expressed or implied, of the Combat Capabilities Development Command Army Research Laboratory or the U.S. Government. The U.S. Government is authorized to reproduce and distribute reprints for Government purposes notwithstanding any copyright notation here on.
\else 
\texttt{ANONYMIZED FOR SUBMISSION}
\fi

\bibliographystyle{plain}
\bibliography{references}

\ifmilcomfinal
\else
\newpage
\appendix
\label{appendix}
\subsection{Experimental Setup}

All experiments were implemented in Pytorch 1.5 \cite{paszke2019pytorch} with Adversarial Attacks using the Adversarial Robustness Toolbox \cite{art2018} on an Intel Xeon E5-2360 with NVIDIA GeForce RTX 2080 Ti and evaluated on the following datasets:
\begin{itemize}
 \item \textbf{MNIST \cite{lecun-mnisthandwrittendigit-2010}:} a 60,000 28 x 28 dataset of digits (0-9).
\item \textbf{CIFAR10 \cite{krizhevsky2009learning}:}  a 50,000 32 x 32 x 3 dataset of images comprised of a size 10 subset of CIFAR100 classes.
\end{itemize}

\subsection{Training Parameters}

\begin{table}[ht]
\centering
\begin{tabular}{|c|c|c|}
\hline
\textbf{Dataset} & \textbf{MNIST} & \textbf{CIFAR10} \\ \hline
\textbf{Model}  & Small Conv. &  Resnet20  \cite{he2016deep}\\ \hline
\textbf{Layers} &  6 & 20 \\ \hline
\textbf{Epochs} & 1 & 25 \\ \hline
\textbf{Batch size} & 64 & 64 \\ \hline
\textbf{Optimizer} & Adam \cite{kingma2014adam} & Adam \cite{kingma2014adam}\\ \hline
\textbf{Learning Rate} & 0.001 & 0.01 \\ \hline
\end{tabular}
\end{table}

\subsection{Attack Parameters}

Attacks were implemented in the Adversarial Robustness Toolbox (ART). If the parameter is not specified, the default value (as specified by ART) should be assumed. All attacks were targeted and targets were chosen at random.

\begin{table}[ht]
\centering
\begin{tabular}{|c|c|c|}
\hline
\multicolumn{3}{|c|}{\textbf{CW (L2)}} \\ \hline
\textbf{Parameter} & \textbf{Description} & \textbf{Value} \\ \hline
\textbf{\begin{tabular}[c]{@{}c@{}}Confidence\\ (float)\end{tabular}} & \begin{tabular}[c]{@{}c@{}}A higher value produces adversarial \\examples that are farther away \\from the original input but \\classified with higher confidence\\ as the target class. \end{tabular} & 0.5\\ \hline
\textbf{\begin{tabular}[c]{@{}c@{}}max\_iter \\(int)\end{tabular}}& Maximum number of iterations. & 5 \\ \hline
\end{tabular}
\caption{Parameters used for the Carlini-Wagner (CW) attack.}
\end{table}

\begin{table}[ht]
\centering
\begin{tabular}{|c|c|c|}
\hline
\multicolumn{3}{|c|}{\textbf{PGD (LINF)}} \\ \hline
\textbf{Parameter} & \multicolumn{1}{c|}{\textbf{Description}} & \textbf{Value} \\ \hline
\textbf{\begin{tabular}[c]{@{}c@{}}eps \\(float)\end{tabular}}  & \begin{tabular}[c]{@{}c@{}}Maximum perturbation that the \\attacker can introduce.\end{tabular} & 0.3 \\ \hline
\textbf{\begin{tabular}[c]{@{}c@{}}max\_iter \\ (int)\end{tabular}} & Maximum number of iterations. & 5 \\ \hline
\textbf{\begin{tabular}[c]{@{}c@{}}norm \\(int)\end{tabular}} & \begin{tabular}[c]{@{}c@{}}The norm of the adversarial \\perturbation. \\Possible values: np.inf, 1, or 2. \end{tabular} & np.inf \\ \hline
\end{tabular}
\caption{Parameters used for the Projected Gradient Descent (PGC) attack.}
\end{table}

\begin{table}[ht]
\centering
\begin{tabular}{|c|c|c|}
\hline
\multicolumn{3}{|c|}{\textbf{JSMA}} \\ \hline
\textbf{Parameter}     & \multicolumn{1}{c|}{\textbf{Description}}& \textbf{Value} \\ \hline
\textbf{theta (float)} & \begin{tabular}[c]{@{}c@{}}Amount of perturbation \\introduced to each modified \\ feature at each step \\ (can be positive or negative).\end{tabular} & 0.3\\ \hline
\end{tabular}
\caption{Parameters used for the Jacobian-based Saliency Map Attack (JSMA).}
\end{table}

\fi

\end{document}